\documentclass{article}

 \usepackage[final]{neurips_2020}

\usepackage[utf8]{inputenc} % allow utf-8 input
\usepackage[T1]{fontenc}    % use 8-bit T1 fonts
\usepackage{hyperref}       % hyperlinks
\usepackage{url}            % simple URL typesetting
\usepackage{booktabs}       % professional-quality tables
\usepackage{amsfonts}       % blackboard math symbols
\usepackage{nicefrac}       % compact symbols for 1/2, etc.
\usepackage{microtype}      % microtypography
\usepackage{algorithm}
\usepackage{algorithmic}
\usepackage{amsmath}
\usepackage{graphicx}
\usepackage{xcolor}
\usepackage{soul}

\title{Learning to Prove Theorems by Learning to Generate Theorems}

\author{%
  Mingzhe Wang \\
  Princeton University\\
  \texttt{mingzhew@cs.princeton.edu} \\
   \And
   Jia Deng \\
   Princeton University \\
   \texttt{jiadeng@cs.princeton.edu} \\
}

\begin{document}

\maketitle

\begin{abstract}
We consider the task of automated theorem proving, a key AI task. Deep learning has shown promise for training theorem provers, but there are limited human-written theorems and proofs available for supervised learning. To address this limitation, we propose to learn a neural generator that automatically synthesizes theorems and proofs for the purpose of training a theorem prover. Experiments on real-world  tasks demonstrate that synthetic data from our approach  improves the theorem prover and advances the state of the art of automated theorem proving in Metamath. Code is available at \href{https://github.com/princeton-vl/MetaGen}{https://github.com/princeton-vl/MetaGen}.
\end{abstract}

\section{Introduction}
\label{intro}

Automated theorem proving aims to automatically
generate a proof given a conjecture (the target theorem) and a knowledge base of known facts, all expressed in a formal language. 
Automated theorem proving is useful
in a wide range of applications, including the verification and synthesis of software and hardware systems~\citep{gu2016certikos,darvas2005theorem,kern1999formal}.

Automated theorem proving boils down to a search problem: finding the sequence of symbol manipulations 
that generate a valid proof. 
The fundamental challenge lies in the explosion of search space, in particular with long proofs
and large knowledge bases. The success of theorem proving thus relies on effective heuristics that guide the prover by deciding the next step the prover should take. 

Deep learning has emerged as a promising approach to learning search heuristics in an automated theorem prover~\citep{irving2016deepmath,whalen2016holophrasm,loos2017deep,bansal2019holist,lee2019mathematical}. 
The search process fundamentally reduces to a sequence of actions on manipulating a set of symbols. Thus a deep network can be
trained to select the best action at each step. 

A key challenge is how to train such networks. Prior work has used human-written theorems and proofs to perform imitation learning and has shown promising results~\citep{loos2017deep,yang2019coqgym,whalen2016holophrasm,paliwal2019graph}. The training data consists of theorems and proofs manually written by human experts in a formal language, and the prover is trained to imitate the proof steps demonstrated by humans. 

However, relying on human-written data has a major drawback: such data has limited availability and scalability. Writing theorems and proofs in a formal language requires highly specialized knowledge and skills, including mathematics, computer programming, and proficiency in the particular formal language. For a CS graduate student, it can take months to master a new formal language such as Mizar, Metamath or HOLight~\citep{wiedijk2003formal}, after which it can take days to formalize a single page of a math textbook. This makes it impractical to crowdsource human-written proofs at large scale. 

\begin{figure}[t]
\centering 
\includegraphics[width=\textwidth]{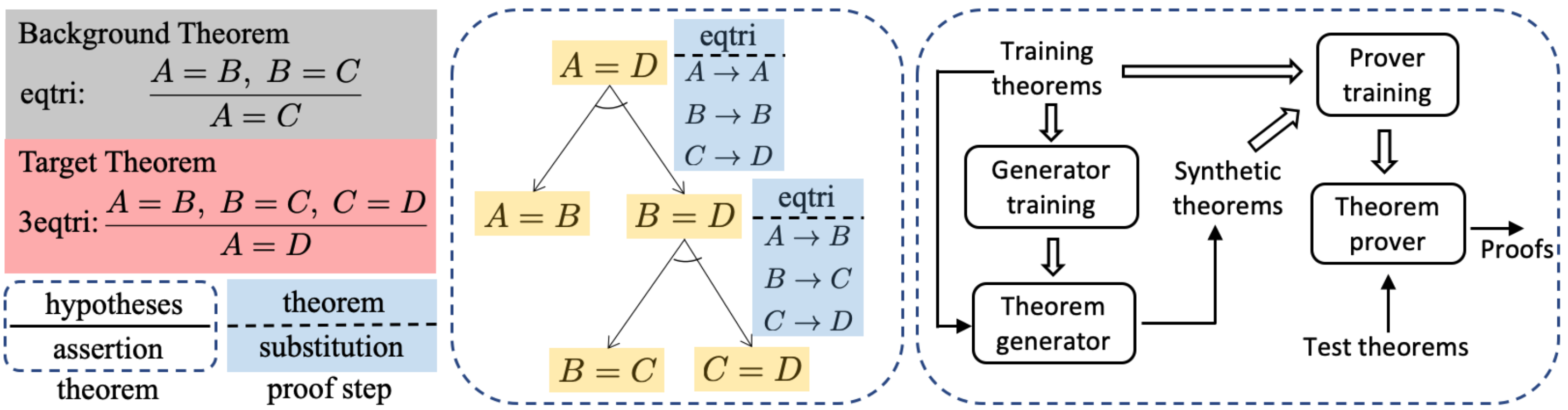}
\vskip -0.1in
\caption{
Left: A proof task. 
Middle: The proof tree of the theorem 3eqtri. 
Each leaf node is a hypothesis and each internal node corresponds to a proof step. 
Right: The overview of our approach. 
}
\label{figure:proof}
\vskip -0.2in
\end{figure}

In this paper, we propose to train a theorem prover using synthetic data. The basic idea is to construct a \emph{generator} that automatically synthesizes new theorems and their proofs, which serve to augment human-written data for training the prover. 

To generate a new theorem and its proof, the generator performs a sequence of symbol manipulations, similar to a prover. It repeatedly applies inference rules on a set of existing theorems and combines their proofs to form the proof of the new theorem.
It is important to note that despite the similarity of operations, the generator has a much easier task than the prover. The generator just needs to generate  \emph{some} new theorem of \emph{its own choice}, whereas the prover needs to find the proof for a particular target theorem specified by someone else. 

One challenge of generating synthetic theorems is that there are infinitely many possibilities but the prover can only use a finite amount of them during training. Not all theorems are equally useful as training data. Thus a key question is how to generate synthetic theorems that are more useful. To this end we make the generator learnable by parameterizing it with deep networks. 

We hypothesize that the generated data will be more useful if they are similar to human-written data. Therefore we use human-written data to train a generator. We consider two scenarios. If the human-written data consist of both theorem statements and their proofs, we train the generator to follow the proof steps in the forward direction, so that a well-trained generator would derive theorems humans tend to derive. If the human-written data consist of only theorem statements but not their proofs, i.e.\@ no human actions to imitate, we use reinforcement learning to let the generator discover good actions that lead to synthetic theorems that are similar to the human-written theorems. To measure similarity between synthetic theorems and human theorems, we use a discriminator trained to distinguish the human theorems from synthetic ones, similar to GANs~\citep{goodfellow2014generative}. 

We instantiate our approach in Metamath~\citep{metamath}, a popular language for formal mathematics, and with Holophrasm~\citep{whalen2016holophrasm}, a Metamath neural prover. We propose a neural theorem generator called ``MetaGen'', which synthesizes new theorems and their proofs expressed in the formalism of Metamath. 
To the best of our knowledge, MetaGen is the first neural generator of synthetic training data for theorem proving. 
Experiments on real-world Metamath tasks show that synthetic data from MetaGen can help train better provers, advancing the state of art in theorem proving on Metamath. 

\section{Related Work}
\noindent\textbf{Automated theorem proving}
Our work is related to prior work on learning to prove theorems~\citep{whalen2016holophrasm,gauthier2018tactictoe,bansal2019holist,yang2019coqgym,loos2017deep,balunovic2018learning,kaliszyk2018reinforcement,bansal2019learning,polu2020generative}.
Our work directly builds off of Holophrasm~\citep{whalen2016holophrasm}, 
a neural-augmented theorem prover for Metamath.
It contains three deep networks 
to generate actions and initial values
to guide proof search following the UCT algorithm~\citep{kocsis2006bandit}. 
\citet{polu2020generative} also build the theorem prover for Metamath by adopting the GPT-like network architectures and pretraining methods and generating proof steps autoregressively. 

TacticToe~\citep{gauthier2018tactictoe}, DeepHOL~\citep{bansal2019holist} and ASTactic~\citep{yang2019coqgym} 
are learning-based theorem provers based on  interactive theorem provers
HOL4~\citep{slind2008brief}, HOL Light~\citep{harrison-hollight} and Coq~\citep{bertot2004coq} respectively. 
\citet{paliwal2019graph} improves DeepHOL by representing formulas as graphs.
\citet{loos2017deep} proposes to learn clause selection by deep learning inside the first-order logic prover E~\citep{schulz2002brainiac}. 

All of these methods are orthogonal to our approach because all of their provers are learned from human-written training data, whereas our prover is trained from human data augmented with synthetic data. Our contribution is on the generation of synthetic data and using such data to train a prover. 

\citet{kaliszyk2018reinforcement,bansal2019holist,bansal2019learning, balunovic2018learning} use reinforcement learning to train provers with  only human-written theorems or SMT conjectures but not proofs. 
During training, a prover collects rewards only upon finding full proofs. 
In contrast, we always train our prover using imitation learning. 
Under the same setting with only human-written theorems but not proofs, we use reinforcement learning to train our generator, whose reward is the similarity between a generated theorem and human-written theorems, as measured by an adversarial discriminator. Our reinforcement learning task is much easier because the reward is continuous and there are many ways to generate theorems similar to human-written ones. 

\noindent\textbf{Synthetic theorem generation} \citet{zombori2019towards, fawzi2019learning} construct theorem provers by training on randomly generated synthetic theorems and evaluate the learned prover on synthetic theorems.
The main difference of our approach is that our generator is optimized through learning, as opposed to random generation. 

\citet{kaliszyk2018reinforcement,jakubuuv2019hammering,urban2008malarea,kaliszyk2014machine,piotrowski2018atpboost} train  theorem provers iteratively. They repeatedly apply the trained prover on existing human theorems and generate new machine proofs to train the prover further. 
In these methods, only new proofs are synthesized and the synthetic proofs are only for existing human theorems; no new theorems are synthesized. In contrast, our approach synthesizes both new theorems and new proofs which could cover a much larger space of possible derivations than the proofs of existing human theorems.

\citet{urban2004mptp,kaliszyk2015learning,kaliszyk2015lemmatization} extract proof tasks from the proofs of human-written theorems, %in MML or ITP theorems, 
such as the intermediate inference steps or their variants. That is, they extract "sub-proofs" from existing proofs. In contrast, we generate entirely new theorems and new proofs that are not part of any existing proofs.

Our work is also related to the line of work on conjecturing~\citep{chvalovskyfirst,urban2020first,colton2012automated}, which aims to generate mathematical conjectures automatically. The generated conjectures are not necessarily true, and their proofs are not required. In contrast, each of our synthetic theorem is guaranteed to be correct and its proof is automatically available. 

\noindent\textbf{Automatic goal generation by self-play}
Our work is similar to the line of work in reinforcement learning
~\citep{pmlr-v80-florensa18a,sukhbaatar2017intrinsic,sukhbaatar2018learning,durugkar2018adversarial}
that deploys two agents in adversary self-play, where one agent to generate tasks for another agent to accomplish. 
We pursue similar ideas in the new context of theorem proving
by learning to generate synthetic theorems to train the prover. Also of note is that we have no adversarial self-play. The goal of the generator is to discover novel theorems similar to human-written ones, not to beat the prover. 

Recently, \citet{huang@learntoprove} introduced a two-player game which
encourages players to learn to 
predict the consistency of logical formulas by self-play.
These two players behave symmetrically and compete with each other in the game. 
In contrast, our generator and prover execute different tasks, and are co-operative. In addition, their game remains a theoretical proposal without any empirical validation, whereas we have performed experiments on 
large-scale data. 

\section{Background on Metamath}
\label{sec:metamath}

Metamath is a language for developing formal mathematics. 
It is one of the simplest formal systems. It has only one inference rule, called \emph{substitution},
but is universally applicable in formalizing a large portion of mathematics
\footnote{Its largest knowledge base, \texttt{set.mm} ranks 3rd 
in the "Formalizing 100 Theorems" challenge~\citep{formalize100}.}
and different types of logic~\citep{metamath}.

\noindent\textbf{Expression and theorem}
A basic building block of Metamath is expressions. An expression is 
a sequence of tokens that follows a set of grammar rules called ``generating axioms''. 
A token is either a constant or a variable.  For example, 
$x+2*y=y+(x+y)$ is an expression, where $x$ and $y$ are two variables.  Each expression corresponds to a unique parse tree where each internal node represents a generating axiom and each leaf node is a token.  

A theorem consists of a set of expressions, one expression as its \emph{assertion} and zero or more expressions as its \emph{hypotheses}. 
The theorem can be understood to state that the hypotheses (e.g.\@ $x^2=1$ and $x>0$) 
entail the assertion (e.g.\@ $x=1$).
Some examples of theorems are shown in Figure~\ref{figure:proof}.

\noindent\textbf{Substitution}
The only inference rule in Metamath is \emph{substitution}, which transforms one expression by replacing each variable with a non-empty new expression. For example, the expression $A=B$ can be transformed to $E+F=C*D$ by the substitution $A\rightarrow E+F$ and $B\rightarrow C*D$. 

Given two expressions $a$ and $b$, we say $b$ can \emph{reach} $a$ or $a$ is \emph{reachable} from $b$ if there exists a substitution that transforms $b$ to $a$. This is equivalent to saying that the parse tree of $b$ can be obtained by ``trimming'' the parse tree of $a$---repeatedly picking an internal node, removing all its descendants, and replacing it with a variable node. Reachability can be checked by comparing parse trees; an algorithm is described in Appendix B. 

\noindent\textbf{Proof step}
A \emph{proof step} is the basic unit of reasoning. 
A proof step in Metamath has two parts: (1) a theorem and (2) a \emph{substitution} that maps each variable in the theorem to a new expression. 
A proof step serves to establish entailment between expressions based on the invoked theorem. 
 For example, let $t$ be the theorem \texttt{over1i}, 
 with the hypothesis $A=B$ and the assertion $(A\;F\;C)=(B\;F\;C)$,
where $\{A,B,C,F\}$ is the set of variables in $t$.
Let $\phi$ be a substitution that maps each variable in $t$ to  a new expression:
$A \rightarrow 2$, $B \rightarrow (1+1)$, $C \rightarrow 2$ and $F \rightarrow + $.
By replacing variables in $t$ with their corresponding expressions given by $\phi$,
we have a new hypothesis $2=(1+1)$ and a new assertion $(2+2)=((1+1)+2)$
This proof step $(t,\phi)$ establishes that  the new hypothesis $2=(1+1)$ entails the new assertion $(2+2)=((1+1)+2)$ based on theorem $t$. 
The new assertion is called the conclusion
and the new hypothesis is called the precondition.
Because a theorem has one assertion and zero or more hypotheses, a proof step thus has one conclusion and zero or more preconditions. 

\noindent\textbf{Proof}
A theorem is proved if we can construct a \emph{proof tree} that connects the hypotheses of the theorem to its assertion through entailment. 
The root node of a proof tree is the assertion of the theorem. Each leaf node of the tree is either a hypothesis of the theorem or empty. 
Each internal node of the tree is an expression and is associated with a proof step that uses an pre-existing theorem, together with an appropriate substitution, to establish the entailment of this internal node by its child nodes. 
Note that if an internal node has an empty child, it means that the proof step has no preconditions. An example proof tree is shown in Figure~\ref{figure:proof}. 

A \emph{proof} is a sequence of proof steps that can be obtained by traversing a proof tree in pre-order. This linearized proof is an equivalent to the tree representation. In this work we will use ``proof'' and ``proof tree'' interchangeably. 

\noindent\textbf{Corpus}
A \emph{corpus} consists of a set of axioms and a sequence of theorems and their corresponding proofs. The proof of each theorem uses only the axioms and the preceding theorems. 

\section{Approach}
\noindent\textbf{Task setup}
We use the standard theorem proving setup in prior work~\cite{irving2016deepmath,bansal2019holist, whalen2016holophrasm}. A \emph{proof task} consists of a \emph{target theorem} (or ``target'' in short) to be proved and a set of \emph{background theorems} to be used as known facts. 
For each theorem in a corpus, we construct a proof task using the theorem as the target theorem and all preceding theorems (i.e. the theorems that humans had available when they were proving the target theorem) as the background theorems. In other words, each theorem in the corpus corresponds to a unique proof task that uses the theorem as the target.  We randomly split all theorems into three disjoint sets: a training set, a validation set, and a test set. Accordingly, we have three corresponding sets of proof tasks using the theorems as targets. More details about this setup in Appendix A.   

\subsection{Generator}
\label{sec:gen}

We propose \emph{MetaGen}, a neural generator that performs forward reasoning to  synthesize theorems. 
It takes a set of training proof tasks as input and outputs a set of synthetic theorems. These synthetic theorems are then combined with original training proof tasks to train the theorem prover (as shown in the right of Fig.~\ref{figure:proof}). The basic operation is generating a proof step---selecting an existing theorem and constructing a substitution. From this single proof step we can derive a new theorem. Now, we can treat this new theorem as an existing theorem and repeat to generate additional new theorems. 

One issue requiring special handling is avoiding generating ``meaningless'' theorems. A meaningless theorem is one that includes falsehood in its hypotheses---as a result it is always provable regardless what the assertion says. It is possible to generate such a theorem if we allow arbitrary substitutions in constructing a proof step. For example, the hypothesis $A=B$ can be substituted into 
$1=2$. Such theorems are valid but unlikely to be useful as training data. 

To avoid meaningless theorems, in constructing a proof step, we require that each new hypothesis produced by substitution must be identical to the full expression of a node in an existing proof tree  (either the root, a leaf, or an internal node), such as the five expressions in yellow boxes in Fig.~\ref{figure:proof}. This prevents introducing false expressions as hypotheses, provided that the existing proofs have no false expressions. See Appendix D about more discussion on meaningless theorems 

A second issue is generating new theorems with multi-step proofs. A single proof step gives a shallow tree. 
To generate theorems with longer proofs, we ``graft'' this shallow tree with existing proof trees or subtrees. For a leaf node $e$ of the shallow tree,  we can replace it with an existing proof tree (or subtree) whose root node is also $e$. For example, suppose the shallow tree proves that $x^2=1$ and $x>0$ entail $x=1$, and there already exists another tree proving that $x^3>0$ entails $x>0$. Then we can join the two trees to generate a new tree proving that $x^3>0$ and $x^2=1$ entail $x=1$. 

To generate theorems and proofs more similar to human-written ones, we impose an additional constraint that a synthesized proof step can only invoke a theorem that has appeared as a background theorem in a training proof task. This is because in the ground-truth proof for a proof task, only the background theorems are invoked in proof steps.  This means that we do not invoke any synthesized theorems. To implement this constraint, the generator constructs proof steps using a restricted set of ``invocable'' theorems pre-specified as input to the generator. 

\noindent\textbf{Initializing existing proof trees}
The generator takes as input a set $E$ of existing theorems and optionally their proof trees, and a set $I$ of invocable theorems, where $E$ and $I$ are the union of the target and background theorems of the training proof tasks respectively.
To enable tree grafting,
it first builds a set $G$ of existing proof trees.
For every theorem in $E$, if its proof tree is available, for every node $e$ in its proof tree, we add to $G$ the subtree that is rooted at $e$ and contains all nodes below $e$. Otherwise, we add to $G$ every hypothesis of this theorem as a single node proof tree. 

Two proof trees are considered equivalent if they have the same root node and the same 
leaf nodes, i.e.\@ they prove the same theorem. Among equivalent trees, we
only keep the smallest one. As a result, $G$ contains all sub-proof trees from all the existing theorems that can be grafted to a new proof step. 

\noindent\textbf{Generating new theorems}
To generate a new theorem,
the key procedure is to construct a proof step and a set $S$ of existing proof trees such that $S$ is a subset of $G$ and 
each precondition of this proof step matches the root node of a proof tree in $S$. This is achieved in three steps as follows: 

\begin{enumerate}
	\item Pick an invocable theorem $t \in I$ according to the frequencies of invocable theorems being used in the proofs of the existing theorems.  
	\item Initialize the set $S$ of proof trees as empty. Initialize the substitution $\phi$ for $t$ as empty. For each hypothesis $h$ of theorem $t$, apply the current substitution $\phi$ to hypothesis $h$ to obtain the transformed expression $h(\phi)$, find all \emph{compatible} proof trees, those whose root nodes are reachable from $h(\phi)$---$h(\phi)$ can be transformed to the root nodes by substitution, which can be determined by comparing parse trees---and perform the following:
	\begin{itemize}
	    \item Select a compatible proof tree $c$ using a \emph{relevance network} (to be described later). For each variable that has not been substituted in $h$, update $\phi$ by assigning the variable a substitute expression to match the root of $c$. Add tree $c$ to set $S$. 
	\end{itemize}
	If no compatible proof tree exists, go to Step 1 and rebuild this proof step from scratch. 
	\item
	If a variable appears in a hypothesis of $t$, 
	its substitution has been determined by matching this hypothesis with the root of a compatible proof tree. 
	For the remaining variables that appear exclusively in the assertion of $t$, use a \emph{subtitution network} (to be described later) to generate substitute expressions for them. 
\end{enumerate}

This proof step gives a one-step proof tree, which we expand to a multi-step proof tree by grafting the trees in set $S$ onto its leaves. This multi-step proof tree is added to $G$ for subsequent generation.
We repeat this procedure to get a set of synthetic theorems 
(pseudo-code in Appendix C).

\noindent\textbf{Relevance network of generator}
The \emph{relevance network} in step 2 is a deep network trained to pick a proof tree from a set of candidates by scoring and ranking them. It uses the same design as the relevance network in Holophrasm~\cite{whalen2016holophrasm} (see Sec.~\ref{sec:prover}) but has different inputs and purposes. 
It takes two sequences of tokens as input.
One input sequence represents the root and leaf nodes of 
a proof tree.
The other sequence consists of two parts.
One part represents
the leaf nodes of the proof trees that have been selected for preceding hypotheses (the hypotheses are processed one by one).
The other part represents the assertion and hypotheses of the invocable theorem transformed by the current substitution, except for the current hypothesis to be processed which is represented by a special token. 
Two GRU encoders convert each input sequence to an embedding vector,
followed by a bilinear layer to output a score 
from the two vectors.
In practice, we limit the number of candidate trees to 2000 for tractability. 

\noindent\textbf{Substitution network of generator}
The substitution network generates the substitution for a target variable
of an invocable theorem. It uses the same design as the ``generation network'' in Holophrasm~\cite{whalen2016holophrasm} (see Sec.~\ref{sec:prover}) but has different inputs and purposes. 
It is a sequence-to-sequence model with the encoder-decoder GRU  network.
It takes as input the sequence of tokens that represents the assertion of the invocable theorem 
and the leaf nodes of the existing proof trees that have been selected to construct a proof step.
The target variable
is represented by a special token. 
The network outputs a sequence of tokens, sampled one by one based on the softmax probabilities. 
 
\noindent\textbf{Generator training}
\label{section:train generator}
We propose two strategies to train
the relevance network and the substitution network,
depending on the availability of human-written proofs. 

Our generator can work without  learnable parameters if we remove
the two deep network and sample new proof steps by randomly picking existing proof trees 
and generating substitutions.
We call such a generator as \emph{MetaGen-Rand}.

Given human-written proofs,
we train \textit{MetaGen-IL} by imitation learning.
Given a proof step $(t,\phi)$ in a human-written proof tree $\mathbf{s}$,
each transformed hypothesis $h(\phi)$ of theorem $t$
 is an internal node of tree $\mathbf{s}$ and is the root of a subtree; we train the relevance network to imitate this step by selecting this subtree among a large set of candidates. 

For a variable $f$ that appears in the assertion but not the hypotheses of $t$, 
the substitution network is trained to produce 
its human-written substitute expression $\phi(f)$.

In the case of only human-written theorems but not their proofs, we can no longer perform imitation learning. We instead use reinforcement learning. The objective is to learn actions to maximize the similarity between the generated theorems and human-written theorems.  
We propose two reward functions to evaluate a generated theorem and  update the two deep networks 
toward the higher rewards via the Reinforce algorithm~\cite{williams1992simple}.
The first reward function is the cross-entropy of a generated theorem given by a language model trained from the human-written theorems. 
The generator from this reward is called \emph{MetaGen-RL-LM}.

The second reward function is given by an adversarial loss similar to GAN~\citep{goodfellow2014generative}---a binary classifier to distinguish the human-written theorems from the generated ones. 
It is pretrained to separate human-written theorems from the theorems generated by \emph{MetaGen-Rand},
and then updated on-the-fly to separate human-written theorems from the theorems 
generated by the current generator. The generator is updated to minimize the adversarial loss. 
We call this generator \emph{MetaGen-RL-Adv}.

More details about the deep networks of the generator are presented in Appendix F.1.

\subsection{Prover}
\label{sec:prover}
We use Holophrasm~\citep{whalen2016holophrasm} as our theorem prover and augment its training with synthetic data. 
Given a proof task,
Holophrasm conducts backward reasoning to prove the target theorem as described in Appendix E. For completeness we briefly summarize how Holophrasm works and refer the reader to~\citet{whalen2016holophrasm} and Appendix E for more details. 

Holophrasm uses Monte Carlo Tree Search (MCTS) to explore multiple branches of actions to find a proof tree.  It involves three learnable deep networks: a \emph{payoff network} to determine which branch is more promising, a \emph{relevance network} to pick a background theorem to construct a proof step, and a \emph{substitution network}\footnote{called the generation network in~\citet{whalen2016holophrasm} but renamed here  
to avoid confusion with the generator.} to generate substitutions.

\subsection{Applicability to other formal systems}
As is standard in related work~\cite{loos2017deep, irving2016deepmath, kaliszyk2018reinforcement, yang2019coqgym}, we instantiate and validate our approach on a single formal system, but our approach
 is applicable to other formal systems such as HOL Light, Coq and Isabelle. 

Our approach can be applied to a new system under the following conditions: (1) the search heuristics of the theorem prover can be trained by imitating ground truth proofs;
(2) the proof of a theorem is a tree of intermediate goals, 
and a proof steps demonstrate the entailment of a goal by its children;
(3) an intermediate goal in the proof is equivalent to a legal theorem.
These conditions are satisfied by the formal systems mentioned above.

To adapt our approach to a new system, 
the main effort is to rewrite 
the procedure of sampling proof steps,
by replacing substitution with 
 inference rules of the new system. 
HOL Light, Coq and Isabelle only provide
tactics as inference rules to decompose a goal into subgoals for backward reasoning.
However, to generate new theorems, we need to execute the corresponding reverse tactics, which are unavailable in their ML environments. 
We leave the experiments on these systems as future work.

\begin{table*}[t]
\caption{Performance of the relevance network of the prover on validation data of \texttt{iset.mm} (top two rows) and \texttt{set.mm} (starting from the third row).
}
\label{table:relevance}
%\vskip 0.1in
\begin{center}
\begin{small}
\begin{sc}
\begin{tabular}{c c c c c c c c}
\toprule
\multicolumn{1}{c}{\bf Human  } 
&\multicolumn{1}{c}{\bf Synthetic  } 
&\multicolumn{1}{c}{\bf Generator}
&\multicolumn{1}{c}{\bf Model}
&\multicolumn{1}{c}{\bf Top-1}
&\multicolumn{1}{c}{\bf Top-5}
&\multicolumn{1}{c}{\bf Top-20}
&\multicolumn{1}{c}{\bf MRR}
%&\multicolumn{1}{c}{\bf Prob}
\\ 
\multicolumn{1}{c}{\bf  proofs } 
&\multicolumn{1}{c}{\bf  proofs } 
&\multicolumn{1}{c}{\bf }
&\multicolumn{1}{c}{\bf }
&\multicolumn{1}{c}{\bf }
&\multicolumn{1}{c}{\bf }
&\multicolumn{1}{c}{\bf }
%&\multicolumn{1}{c}{\bf }
&\multicolumn{1}{c}{\bf }\\
\midrule
7123 (iset) & 0 & - & relevance & 43.27 & 69.57 & 89.68 & 0.5535 \\
7123 (iset) & 1M & \textit{MetaGen-IL} & relevance & 45.10 & 71.00 & 89.46 & 0.5699
\\ \midrule
0 & 0&- & tf-idf & 14.28 & 21.13 & 32.55 &0.1877 %& 0.0062 
\\
0 & 0&- & relevance & 0.96 & 5.33 & 15.67 & 0.0445 %& 0.0068 
\\
0 & 300K &\textit{MetaGen-Rand}& relevance & 24.22 & 37.27 & 49.92 & 0.3093 %& 0.2272   
\\
0 & 300K &\textit{MetaGen-RL-LM}& relevance & 24.74 & 37.66 & 54.22 & 0.3182 %& 0.2315 
\\ 
0 & 300K &\textit{MetaGen-RL-Adv}& relevance & 25.07 & 39.33 & 50.23 & 0.3242 %& 0.2426 
\\\midrule
2179 (10\%) & 0 & -&  relevance& 41.24 & 67.56 & 86.84 & 0.5356 %& 0.3732 
\\
2179 (10\%) & 1M&\textit{MetaGen-Rand}& relevance& 45.44 & 70.13 & 88.33 & 0.5692 %& 0.3768  
\\
2179 (10\%) & 1M&\textit{MetaGen-IL}& relevance& 46.10 & 71.12 & 89.38 & 0.5772 %& 0.3863 
\\ 
4358 (20\%) & 0 & - & relevance & 47.02 & 72.45 & 89.48 & 0.5870 \\
\midrule
21786 (100\%)&0 & -&  relevance& 51.52 & 78.56 & 93.41 & 0.6367 %& 0.4396 
\\
21786 (100\%)& 10M & \textit{MetaGen-Rand}& relevance & 52.08 & 77.76 & 92.83 & 0.6375 %& 0.4387 
\\
21786 (100\%)& 10M & \textit{MetaGen-IL}& relevance & 53.20 & 78.73 & 93.13 & 0.6474 %& 0.4530 
\\
\bottomrule
\end{tabular}
\end{sc}
\end{small}
\end{center}
\vskip -0.1in
\end{table*}

\begin{table*}[t]
\caption{Performance of the substitution network of the prover on validation data of \texttt{iset.mm} (top two rows) and \texttt{set.mm} (starting from the third row).}
\label{table:generation}
\begin{center}
\begin{small}
\begin{sc}
\begin{tabular}{c c c c c c}
\toprule
\multicolumn{1}{c}{\bf Human proofs } 
&\multicolumn{1}{c}{\bf Synthetic proofs } 
&\multicolumn{1}{c}{\bf Generator}
&\multicolumn{1}{c}{\bf Model}
&\multicolumn{1}{c}{\bf Prob}
&\multicolumn{1}{c}{\bf Accuracy}
\\ \midrule
% 7125 &0 & -& 43.27 & 69.57 & 89.68 & 0.5535 & 0.1723 & 49.45 & 378 
%\\
%7125 & 1M & \textit{MetaGen-IL} & 45.10 & 71.00 & 89.46 & 0.5699 & 0.2554 & 57.81 & 398
7123 (iset) & 0 & - & substitution & 0.1723 & 49.45 \\
7123 (iset) & 1M & \textit{MetaGen-IL} & substitution & 0.2554 & 57.81 \\
\midrule 
0 & 0&- & lauguage model & 0.0032 & 9.06 \\
0 & 0&- &substitution & 0.0008 & 0.01 \\
0 & 300K &\textit{MetaGen-Rand}& substitution & 0.0103  &  29.68  \\
0 & 300K &\textit{MetaGen-RL-LM}& substitution & 0.0181  &  24.33  \\ 
0 & 300K &\textit{MetaGen-RL-Adv}& substitution & 0.0186  &  31.38  \\ \midrule
2179 (10\%) & 0 & -&  substitution& 0.2738  & 58.91  \\
2179 (10\%) & 1M&\textit{MetaGen-Rand}&substitution & 0.3203 & 61.78 \\
2179 (10\%)& 1M&\textit{MetaGen-IL}& substitution& 0.3710 & 66.56  \\ 
4358 (20\%) & 0 & -&  substitution& 0.3765  & 67.07  \\
\midrule
21786 (100\%)&0 & -&  substitution& 0.6142 &81.57  \\
21786 (100\%)& 10M&\textit{MetaGen-Rand}&substitution & 0.6439 & 81.85 \\
21786 (100\%)& 10M & \textit{MetaGen-IL}& substitution & 0.6847 & 83.90  \\
\bottomrule
\end{tabular}
\end{sc}
\end{small}
\end{center}
\vskip -0.1in
\end{table*}

\section{Experiments}

\noindent\textbf{Dataset}
We experiment on two Metamath knowledge bases:  \texttt{iset.mm} and \texttt{set.mm}. 
\texttt{iset.mm} formalizes intuitionistic logic and 
contains 463 axioms and 8916 theorems, which give rise to 8916 corresponding proof tasks. These proof tasks are divided into 7123 training tasks, 890 validation tasks and 903 test tasks. 
We use the same version of \texttt{set.mm} as~\citet{whalen2016holophrasm}. 
It formalizes the ZFC set theory
and contains 1099 axioms and 27218 theorems, 
which give rise to 27218 corresponding proof tasks. These proof tasks are divided into 21786 training tasks, 2712 validation tasks and 2720 test tasks.

\noindent\textbf{Training protocol}
On \texttt{set.mm}, we control for the number of human proofs provided during training. Specifically, we compare our approach to baselines while including either 0\%, 10\%, or 100\% of the human proofs. We also report the baseline with 20\% human proofs for comparison.

\noindent\textbf{Implementation details}
We train the generator on the training set and use the trained generator to generate synthetic theorems and proofs. The prover is trained on both training and synthetic proofs.

On \texttt{iset.mm}, we generate 1M unique synthetic theorems. 
On \texttt{set.mm}, we generate 300K unique theorems for the setting of 0\% of human proofs (after discarding any duplicates) and 1M unique theorems for 10\% of the human training proofs. We generate 10M theorems for the setting of 100\% of human proofs, by generating 1M unique theorems a time (maximum allowed by memory limit) and repeating 10 times. 

During the training of the relevance network of the prover, 
we filter out the "trivial" proof steps.
A goal is trivial if it is reachable from the assertion of a background theorem $b$ and $b$ has no hypotheses, 
because this goal can be decomposed by $b$ without generating any new subgoals. 
By removing the training proof steps that have trivial goals
when we train the relevance network,
the performance of the prover is improved as shown in Tab.~\ref{table:prover}.

Please refer to 
Appendix F for more details
about the implementation and baselines.

\begin{table*}[t]
\caption{Number of theorems proved on test data of \texttt{iset.mm} (top two rows) and \texttt{set.mm} (starting from the third row).
$\dagger$: without removing the trivial proof steps from the training data of the relevance network.}
\label{table:prover}
\begin{center}
\begin{small}
\begin{sc}
\begin{tabular}{c c c c c}
\toprule
\multicolumn{1}{c}{\bf Human  } 
&\multicolumn{1}{c}{\bf Synthetic  } 
&\multicolumn{1}{c}{\bf Generator}
&\multicolumn{1}{c}{\bf Prover}
&\multicolumn{1}{c}{\bf Test proofs }
\\
\multicolumn{1}{c}{\bf  proofs } 
&\multicolumn{1}{c}{\bf  proofs } 
&\multicolumn{1}{c}{\bf }
&\multicolumn{1}{c}{\bf }
&\multicolumn{1}{c}{\bf found}
\\ \midrule
7123 (iset)& 0 & - & Holophrasm & 378 \\
7123 (iset)& 1M & \textit{MetaGen-IL}& Holophrasm & 398  \\
\midrule
0 & 0&- & tf-idf $\&$ LM & 312 \\
0 & 0&- &Holophrasm & 219 \\
0 & 300K & \textit{MetaGen-Rand} & Holophrasm & 346 \\
0 & 300K &\textit{MetaGen-RL-LM}& Holophrasm & 351  \\ 
0 & 300K &\textit{MetaGen-RL-Adv}& Holophrasm & 357  \\ 
\midrule
2179 (10\%) & 0 & -&  Holophrasm& 454  \\
2179 (10\%) & 1M&\textit{MetaGen-Rand}&Holophrasm & 457 \\
2179 (10\%) & 1M&\textit{MetaGen-IL}& Holophrasm& 472  \\ 
4358 (20\%) & 0 & -&  Holophrasm& 476  \\
\midrule
21786 (100\%)&0 & -& Holophrasm('16) & 388 \\
21786 (100\%)&0 & -&  Holophrasm& 557  \\
21786 (100\%)& 10M & \textit{MetaGen-Rand}& Holophrasm & 565 \\
21786 (100\%)& 10M & \textit{MetaGen-IL}& Holophrasm$^\dagger$ & 574  \\
21786 (100\%)& 10M & \textit{MetaGen-IL}& Holophrasm & 600  \\
\bottomrule
\end{tabular}
\end{sc}
\end{small}
\end{center}
\vskip -0.2in
\end{table*}

\subsection{Results}

To validate the effectiveness of our theorem generator, we evaluate provers trained on the synthetic data and compare them against various baselines. 

\noindent\textbf{Relevance network of prover}
We evaluate how synthetic data can improve the relevance network of Holophrasm. The relevance network assigns a score to each candidate background theorem. 
We use two metrics: (1) top-k accuracy defined as the percentage of times a groundtruth background theorems is ranked in the top k and 
(2) mean reciprocal rank (MRR) of every groundtruth background theorem among all candidates of its corresponding proof step. Both of them are the higher the better.

We evaluate the relevance network combined with different generators. We also evaluate with tf-idf similarity between sequences of tokens. 
In Tab.~\ref{table:relevance},
we see that synthetic data brings significant improvement in all settings and
the best performance is achieved with our trained generators. 

\noindent\textbf{Substitution network of prover}
We evaluate how synthetic data can improve the substitution network of Holophrasm. 
The substitution network predicts the probability of each token at each position under teacher forcing.
We use two metrics: (1) accuracy, defined as the percentage of times the tokens in the groundtruth substitutions have the highest probabilities and (2) the average probability to generate the groundtruth substitutions normalized by its length. Tab.~\ref{table:generation} reports the results, including the result of a language model. 
In all settings, synthetic data brings significant improvement. 
The best performance is achieved with our trained generators. 

\noindent\textbf{Prover}
To evaluate the prover as a whole, we follow the same protocol of~\citet{whalen2016holophrasm} (more details in Appendix F.2) and report the number of theorems proved. We compare with the original Holophrasm prover proposed by~\citet{whalen2016holophrasm} trained 
by imitation learning on human-written proofs only. With zero human-written proofs for prover training, 
we also evaluate TF-IDF \& LM, an ablated version of Holophrasm that needs no training proofs---we remove the relevance network and instead pick a background theorem using tf-idf similarity; we replace the substitution network with a language model of theorem statements.

As shown in Tab.~\ref{table:prover},
the performance of the prover shares the same pattern as the relevance and substitution network.
On both \texttt{iset.mm} and \texttt{set.mm}, the provers trained on synthetic data
consistently prove more theorems 
than the provers trained on human proofs only.
On \texttt{set.mm}, with 10\% human proofs, the use of synthetic proofs almost achieve the same effect by doubling the number of human proofs (472 vs 476 proved theorems). 
The provers trained with learnable generators perform 
better than the provers trained with \textit{MetaGen-Rand}.

Our GPU re-implementation of Holophrasm finds 557 proofs trained on 100\% of human proofs, more than the number reported in~\citet{whalen2016holophrasm}.
We believe this is due to the fact that our prover runs faster on GPUs. 

By removing the trivial proof steps from the training data of the relevance network of the prover, 
the number of proved theorems on the test set increases from 574 to 600.

\citet{polu2020generative}
demonstrate significant improvement on theorem proving of the \texttt{set.mm} benchmark by using very large Transformer~\citep{vaswani2017attention} models. Their model can prove 29.22\% of test theorems (our percentage is 22.06\%). We note a couple potential differences in experimental setup, which may make our results not directly comparable. They appear to use a different version of the \texttt{set.mm} knowledge base which has about 38k proofs (ours has 27218 proofs); their  evaluation protocol may be different (our prover has a time limit of 5 minutes for each run while their time limit is not mentioned). 

Please refer to Appendix G for the examples of synthetic theorems.

\section{Conclusion}
We have proposed a neural generator that automatically synthesizes theorems and proofs for the purpose of training a theorem prover. Experiments on real-world  tasks have demonstrated that synthetic data from our approach improves the theorem prover and advances the state of the art of automated theorem proving in Metamath.

\noindent\textbf{Acknowledgements} This work is partially supported by the National Science Foundation under Grant IIS-1903222 and the Office of Naval Research under Grant N00014-20-1-2634. 

\section*{Broader Impact}
Our work addresses automated theorem proving. 
A successful automated theorem prover can help us write programs that are provably correct, which is essential to safety-critical applications, such as software for autonomous driving. On the other hand, since the correctness of the found proofs and synthesized programs relies on the correctness of the underlying theorem prover, bugs in the prover can lead to catastrophic failure. 

\bibliography{neurips_2020}

\begin{thebibliography}{44}
\providecommand{\natexlab}[1]{#1}
\providecommand{\url}[1]{\texttt{#1}}
\expandafter\ifx\csname urlstyle\endcsname\relax
  \providecommand{\doi}[1]{doi: #1}\else
  \providecommand{\doi}{doi: \begingroup \urlstyle{rm}\Url}\fi

\bibitem[Balunovic et~al.(2018)Balunovic, Bielik, and
  Vechev]{balunovic2018learning}
Balunovic, M., Bielik, P., and Vechev, M.
\newblock Learning to solve smt formulas.
\newblock In \emph{Advances in Neural Information Processing Systems}, pp.\
  10317--10328, 2018.

\bibitem[Bansal et~al.(2019{\natexlab{a}})Bansal, Loos, Rabe, Szegedy, and
  Wilcox]{bansal2019holist}
Bansal, K., Loos, S., Rabe, M., Szegedy, C., and Wilcox, S.
\newblock Holist: An environment for machine learning of higher order logic
  theorem proving.
\newblock In \emph{International Conference on Machine Learning}, pp.\
  454--463, 2019{\natexlab{a}}.

\bibitem[Bansal et~al.(2019{\natexlab{b}})Bansal, Loos, Rabe, and
  Szegedy]{bansal2019learning}
Bansal, K., Loos, S.~M., Rabe, M.~N., and Szegedy, C.
\newblock Learning to reason in large theories without imitation.
\newblock \emph{arXiv preprint arXiv:1905.10501}, 2019{\natexlab{b}}.

\bibitem[Bertot \& Cast{\'e}ran(2004)Bertot and Cast{\'e}ran]{bertot2004coq}
Bertot, Y. and Cast{\'e}ran, P.
\newblock Coq’art: the calculus of inductive constructions, 2004.

\bibitem[Chvalovsk{\`y} et~al.(2019)Chvalovsk{\`y}, Gauthier, and
  Urban]{chvalovskyfirst}
Chvalovsk{\`y}, K., Gauthier, T., and Urban, J.
\newblock First experiments with data driven conjecturing.
\newblock \emph{4th Conference on Artificial Intelligence and Theorem Proving},
  2019.

\bibitem[Colton(2012)]{colton2012automated}
Colton, S.
\newblock \emph{Automated theory formation in pure mathematics}.
\newblock Springer Science \& Business Media, 2012.

\bibitem[Darvas et~al.(2005)Darvas, H{\"a}hnle, and Sands]{darvas2005theorem}
Darvas, {\'A}., H{\"a}hnle, R., and Sands, D.
\newblock A theorem proving approach to analysis of secure information flow.
\newblock In \emph{International Conference on Security in Pervasive
  Computing}, pp.\  193--209. Springer, 2005.

\bibitem[Durugkar \& Stone(2018)Durugkar and Stone]{durugkar2018adversarial}
Durugkar, I. and Stone, P.
\newblock Adversarial goal generation for intrinsic motivation.
\newblock In \emph{Thirty-Second AAAI Conference on Artificial Intelligence},
  2018.

\bibitem[Fawzi et~al.(2019)Fawzi, Malinowski, Fawzi, and
  Fawzi]{fawzi2019learning}
Fawzi, A., Malinowski, M., Fawzi, H., and Fawzi, O.
\newblock Learning dynamic polynomial proofs.
\newblock In \emph{Advances in Neural Information Processing Systems}, pp.\
  4181--4190, 2019.

\bibitem[Florensa et~al.(2018)Florensa, Held, Geng, and
  Abbeel]{pmlr-v80-florensa18a}
Florensa, C., Held, D., Geng, X., and Abbeel, P.
\newblock Automatic goal generation for reinforcement learning agents.
\newblock In Dy, J. and Krause, A. (eds.), \emph{Proceedings of the 35th
  International Conference on Machine Learning}, volume~80 of \emph{Proceedings
  of Machine Learning Research}, pp.\  1515--1528, Stockholmsmässan, Stockholm
  Sweden, 10--15 Jul 2018. PMLR.
\newblock URL \url{http://proceedings.mlr.press/v80/florensa18a.html}.

\bibitem[Gauthier et~al.(2018)Gauthier, Kaliszyk, and
  Urban]{gauthier2018tactictoe}
Gauthier, T., Kaliszyk, C., and Urban, J.
\newblock Tactictoe: Learning to reason with hol4 tactics.
\newblock \emph{arXiv preprint arXiv:1804.00595}, 2018.

\bibitem[Goodfellow et~al.(2014)Goodfellow, Pouget-Abadie, Mirza, Xu,
  Warde-Farley, Ozair, Courville, and Bengio]{goodfellow2014generative}
Goodfellow, I., Pouget-Abadie, J., Mirza, M., Xu, B., Warde-Farley, D., Ozair,
  S., Courville, A., and Bengio, Y.
\newblock Generative adversarial nets.
\newblock In \emph{Advances in neural information processing systems}, pp.\
  2672--2680, 2014.

\bibitem[Gu et~al.(2016)Gu, Shao, Chen, Wu, Kim, Sj{\"o}berg, and
  Costanzo]{gu2016certikos}
Gu, R., Shao, Z., Chen, H., Wu, X.~N., Kim, J., Sj{\"o}berg, V., and Costanzo,
  D.
\newblock Certikos: An extensible architecture for building certified
  concurrent $\{$OS$\}$ kernels.
\newblock In \emph{12th $\{$USENIX$\}$ Symposium on Operating Systems Design
  and Implementation ($\{$OSDI$\}$ 16)}, pp.\  653--669, 2016.

\bibitem[Harrison(2009)]{harrison-hollight}
Harrison, J.
\newblock {HOL} {L}ight: An overview.
\newblock In Berghofer, S., Nipkow, T., Urban, C., and Wenzel, M. (eds.),
  \emph{Proceedings of the 22nd International Conference on Theorem Proving in
  Higher Order Logics, TPHOLs 2009}, volume 5674 of \emph{Lecture Notes in
  Computer Science}, pp.\  60--66, Munich, Germany, 2009. Springer-Verlag.

\bibitem[Huang(2019)]{huang@learntoprove}
Huang, D.
\newblock On learning to prove.
\newblock \emph{arXiv preprint arXiv:1904.11099}, 2019.

\bibitem[Irving et~al.(2016)Irving, Szegedy, Alemi, E{\'e}n, Chollet, and
  Urban]{irving2016deepmath}
Irving, G., Szegedy, C., Alemi, A.~A., E{\'e}n, N., Chollet, F., and Urban, J.
\newblock Deepmath-deep sequence models for premise selection.
\newblock In \emph{Advances in Neural Information Processing Systems}, pp.\
  2235--2243, 2016.

\bibitem[Jakub\r{u}v \& Urban(2019)Jakub\r{u}v and
  Urban]{jakubuuv2019hammering}
Jakub\r{u}v, J. and Urban, J.
\newblock Hammering mizar by learning clause guidance.
\newblock \emph{arXiv preprint arXiv:1904.01677}, 2019.

\bibitem[Kaliszyk \& Urban(2015)Kaliszyk and Urban]{kaliszyk2015learning}
Kaliszyk, C. and Urban, J.
\newblock Learning-assisted theorem proving with millions of lemmas.
\newblock \emph{Journal of symbolic computation}, 69:\penalty0 109--128, 2015.

\bibitem[Kaliszyk et~al.(2014)Kaliszyk, Urban, and
  Vysko{\v{c}}il]{kaliszyk2014machine}
Kaliszyk, C., Urban, J., and Vysko{\v{c}}il, J.
\newblock Machine learner for automated reasoning 0.4 and 0.5.
\newblock \emph{arXiv preprint arXiv:1402.2359}, 2014.

\bibitem[Kaliszyk et~al.(2015)Kaliszyk, Urban, and
  Vysko{\v{c}}il]{kaliszyk2015lemmatization}
Kaliszyk, C., Urban, J., and Vysko{\v{c}}il, J.
\newblock Lemmatization for stronger reasoning in large theories.
\newblock In \emph{International Symposium on Frontiers of Combining Systems},
  pp.\  341--356. Springer, 2015.

\bibitem[Kaliszyk et~al.(2018)Kaliszyk, Urban, Michalewski, and
  Ol{\v{s}}{\'a}k]{kaliszyk2018reinforcement}
Kaliszyk, C., Urban, J., Michalewski, H., and Ol{\v{s}}{\'a}k, M.
\newblock Reinforcement learning of theorem proving.
\newblock In \emph{Advances in Neural Information Processing Systems}, pp.\
  8822--8833, 2018.

\bibitem[Kern \& Greenstreet(1999)Kern and Greenstreet]{kern1999formal}
Kern, C. and Greenstreet, M.~R.
\newblock Formal verification in hardware design: a survey.
\newblock \emph{ACM Transactions on Design Automation of Electronic Systems
  (TODAES)}, 4\penalty0 (2):\penalty0 123--193, 1999.

\bibitem[Kingma \& Ba(2014)Kingma and Ba]{kingma2014adam}
Kingma, D.~P. and Ba, J.
\newblock Adam: A method for stochastic optimization.
\newblock \emph{arXiv preprint arXiv:1412.6980}, 2014.

\bibitem[Kocsis \& Szepesv{\'a}ri(2006)Kocsis and
  Szepesv{\'a}ri]{kocsis2006bandit}
Kocsis, L. and Szepesv{\'a}ri, C.
\newblock Bandit based monte-carlo planning.
\newblock In \emph{European conference on machine learning}, pp.\  282--293.
  Springer, 2006.

\bibitem[Lee et~al.(2019)Lee, Szegedy, Rabe, Loos, and
  Bansal]{lee2019mathematical}
Lee, D., Szegedy, C., Rabe, M.~N., Loos, S.~M., and Bansal, K.
\newblock Mathematical reasoning in latent space.
\newblock \emph{arXiv preprint arXiv:1909.11851}, 2019.

\bibitem[Loos et~al.(2017)Loos, Irving, Szegedy, and Kaliszyk]{loos2017deep}
Loos, S., Irving, G., Szegedy, C., and Kaliszyk, C.
\newblock Deep network guided proof search.
\newblock \emph{arXiv preprint arXiv:1701.06972}, 2017.

\bibitem[Megill \& Wheeler(2019)Megill and Wheeler]{metamath}
Megill, N. and Wheeler, D.
\newblock \emph{Metamath: A Computer Language for Mathematical Proofs}.
\newblock Lulu Press, Morrisville, North Carolina, 2019.
\newblock { http://us.metamath.org/downloads/metamath.pdf}.

\bibitem[Paliwal et~al.(2019)Paliwal, Loos, Rabe, Bansal, and
  Szegedy]{paliwal2019graph}
Paliwal, A., Loos, S., Rabe, M., Bansal, K., and Szegedy, C.
\newblock Graph representations for higher-order logic and theorem proving.
\newblock \emph{arXiv preprint arXiv:1905.10006}, 2019.

\bibitem[Piotrowski \& Urban(2018)Piotrowski and Urban]{piotrowski2018atpboost}
Piotrowski, B. and Urban, J.
\newblock Atpboost: Learning premise selection in binary setting with atp
  feedback.
\newblock In \emph{International Joint Conference on Automated Reasoning}, pp.\
   566--574. Springer, 2018.

\bibitem[Polu \& Sutskever(2020)Polu and Sutskever]{polu2020generative}
Polu, S. and Sutskever, I.
\newblock Generative language modeling for automated theorem proving.
\newblock \emph{arXiv preprint arXiv:2009.03393}, 2020.

\bibitem[Schulz(2002)]{schulz2002brainiac}
Schulz, S.
\newblock E--a brainiac theorem prover.
\newblock \emph{Ai Communications}, 15\penalty0 (2, 3):\penalty0 111--126,
  2002.

\bibitem[Slind \& Norrish(2008)Slind and Norrish]{slind2008brief}
Slind, K. and Norrish, M.
\newblock A brief overview of hol4.
\newblock In \emph{International Conference on Theorem Proving in Higher Order
  Logics}, pp.\  28--32. Springer, 2008.

\bibitem[Sukhbaatar et~al.(2017)Sukhbaatar, Lin, Kostrikov, Synnaeve, Szlam,
  and Fergus]{sukhbaatar2017intrinsic}
Sukhbaatar, S., Lin, Z., Kostrikov, I., Synnaeve, G., Szlam, A., and Fergus, R.
\newblock Intrinsic motivation and automatic curricula via asymmetric
  self-play.
\newblock \emph{arXiv preprint arXiv:1703.05407}, 2017.

\bibitem[Sukhbaatar et~al.(2018)Sukhbaatar, Denton, Szlam, and
  Fergus]{sukhbaatar2018learning}
Sukhbaatar, S., Denton, E., Szlam, A., and Fergus, R.
\newblock Learning goal embeddings via self-play for hierarchical reinforcement
  learning.
\newblock \emph{arXiv preprint arXiv:1811.09083}, 2018.

\bibitem[Urban(2004)]{urban2004mptp}
Urban, J.
\newblock Mptp--motivation, implementation, first experiments.
\newblock \emph{Journal of Automated Reasoning}, 33\penalty0 (3-4):\penalty0
  319--339, 2004.

\bibitem[Urban \& Jakub\r{u}v(2020)Urban and Jakub\r{u}v]{urban2020first}
Urban, J. and Jakub\r{u}v, J.
\newblock First neural conjecturing datasets and experiments.
\newblock \emph{arXiv preprint arXiv:2005.14664}, 2020.

\bibitem[Urban et~al.(2008)Urban, Sutcliffe, Pudl{\'a}k, and
  Vysko{\v{c}}il]{urban2008malarea}
Urban, J., Sutcliffe, G., Pudl{\'a}k, P., and Vysko{\v{c}}il, J.
\newblock Malarea sg1-machine learner for automated reasoning with semantic
  guidance.
\newblock In \emph{International Joint Conference on Automated Reasoning}, pp.\
   441--456. Springer, 2008.

\bibitem[Vaswani et~al.(2017)Vaswani, Shazeer, Parmar, Uszkoreit, Jones, Gomez,
  Kaiser, and Polosukhin]{vaswani2017attention}
Vaswani, A., Shazeer, N., Parmar, N., Uszkoreit, J., Jones, L., Gomez, A.~N.,
  Kaiser, {\L}., and Polosukhin, I.
\newblock Attention is all you need.
\newblock In \emph{Advances in neural information processing systems}, pp.\
  5998--6008, 2017.

\bibitem[Whalen(2016)]{whalen2016holophrasm}
Whalen, D.
\newblock Holophrasm: a neural automated theorem prover for higher-order logic.
\newblock \emph{arXiv preprint arXiv:1608.02644}, 2016.

\bibitem[Wiedijk(2003)]{wiedijk2003formal}
Wiedijk, F.
\newblock Formal proof sketches.
\newblock In \emph{International Workshop on Types for Proofs and Programs},
  pp.\  378--393. Springer, 2003.

\bibitem[Wiedijk(2019)]{formalize100}
Wiedijk, F.
\newblock Formalizing 100 theorems.
\newblock \url{http://www.cs.ru.nl/~freek/100/}, 2019.

\bibitem[Williams(1992)]{williams1992simple}
Williams, R.~J.
\newblock Simple statistical gradient-following algorithms for connectionist
  reinforcement learning.
\newblock \emph{Machine learning}, 8\penalty0 (3-4):\penalty0 229--256, 1992.

\bibitem[Yang \& Deng(2019)Yang and Deng]{yang2019coqgym}
Yang, K. and Deng, J.
\newblock Learning to prove theorems via interacting with proof assistants.
\newblock In \emph{International Conference on Machine Learning}, 2019.

\bibitem[Zombori et~al.(2019)Zombori, Csisz{\'a}rik, Michalewski, Kaliszyk, and
  Urban]{zombori2019towards}
Zombori, Z., Csisz{\'a}rik, A., Michalewski, H., Kaliszyk, C., and Urban, J.
\newblock Towards finding longer proofs.
\newblock \emph{arXiv preprint arXiv:1905.13100}, 2019.

\end{thebibliography}
\bibliographystyle{icml2020}

\section*{Appendix}

\subsection*{A. Task setup}
We use the standard theorem proving setup in prior work~\cite{irving2016deepmath,bansal2019holist, whalen2016holophrasm}.
Suppose we have a sequence of theorems $(t_1, t_2, ..., t_n)$, where each theorem appear at the order it is proved by mathematicians. For each theorem $t_i$, we construct a proof task
that proving $t_i$ (as the target theorem) using all its preceding theorems $(t_1, ..., t_{i-1})$ (as the background theorems), such that the prover has the same set of known facts as mathematicians to prove $t_i$. Then we randomly split the five proof tasks into three sets for training, validation and testing.

It is important to note that a theorem can serve both as a target theorem in the test set and as a background theorem in the training set. This is a standard setup and is not ``training on the test set''---a background theorem is used as a known fact in a training proof task and only its statement is provided, not its proof; seeing the statement of a background theorem during training does not tell us how to prove it during testing. 

\subsection*{B. Checking reachability between expressions}
\label{app:reach}
For an expression $e$, 
let $r_e$ be the root node of the parse tree of $e$. Each node in the parse tree represents either a generating axiom (if internal node) or a token (if leaf node). 
We check if expression $b$ can reach expression $a$ by comparing their parse trees $r_a$
and $r_b$ through the following procedure:
\begin{algorithm}[t]
   \caption{Function \texttt{Reachable}($n_a$, $n_b$, $\phi$)}
   \label{algo:reach}
   \begin{algorithmic}
    \STATE {\bfseries Input:} node $n_a$, node $n_b$, substitution $\phi$ 
    \STATE {\bfseries Output:} \emph{True} if $n_b$ could reach $n_a$, otherwise \emph{False}
    \IF{$n_b$ represents a variable $f$}  
        \IF{$f$ in $\phi$}
            \IF{$\phi(f)=n_a$}
                \STATE{\bfseries return} \emph{True} \COMMENT {Consistent with the current substitution}
            \ELSE 
                \STATE{\bfseries return} \emph{False} \COMMENT{Conflict with a preceding branch}
            \ENDIF
        \ELSE
            \STATE $\phi(f)=n_a$ \COMMENT {Variable $f$ should be replaced by $n_a$}
            \STATE{\bfseries return} \emph{True}
        \ENDIF
    \ELSE
        \IF{$n_a$ and $n_b$ represent the same generating axiom or constant}
            \FOR{$i=1$ {\bfseries to} len$(c_{n_a})$}
                \STATE\COMMENT{$c_n$ is the list of children of node $n$}
                \IF{\texttt{Reachable}$(c_{n_a}[i], c_{n_b}[i], \phi) =$ false}
                    \STATE\COMMENT {A pair of child nodes doesn't match}
                    \STATE{\bfseries return} \emph{False} 
                \ENDIF
            \ENDFOR
            \STATE\COMMENT{Every child of $n_b$ could reach a child of $n_a$}
            \STATE{\bfseries return} \emph{True} 
        \ELSE
            \STATE{\bfseries return} \emph{False} \COMMENT{Two nodes have different values}
        \ENDIF
    \ENDIF
   \end{algorithmic}
   
\end{algorithm}

\begin{enumerate}
    \item Initialize the substitution $\phi$ as empty.
    \item Compare the two root nodes.
    \begin{itemize}
        \item If root node $r_b$ represents a variable $f$, do the following:
            \begin{itemize}
                \item If the substitute expression $\phi(f)$ is not determined, let $\phi(f) \leftarrow r_a$. Return \emph{True} (i.e. reachable). 
                \item If $\phi(f)=r_a$, return \emph{True} (i.e. reachable) because we can replace $f$ with $r_a$.   
                \item Otherwise return \emph{False} (unreachable), because $r_a$ 
                conflicts with the current substitution $\phi$. 
            \end{itemize}
        \item If the two root nodes represent the same generating aixom or constant, 
            repeat Step 2 to check if each child of $r_a$ is reachable from 
            the corresponding child of $r_b$. 
            \begin{itemize}
                \item If every child of $r_a$ is reachable 
                    from the corresponding child of $r_b$, return \emph{True}.
                \item Otherwise return \emph{False}.
            \end{itemize}
        \item Otherwise return \emph{False}, because the two root nodes have different values and they can not be matched.
    \end{itemize}
\end{enumerate}

This procedure is summarized in Algorithm~\ref{algo:reach}.

\subsection*{C. Pseudo-code for MetaGen}
\label{app:code}
Algorithm~\ref{algo:step} summarizes the procedure to construct a proof step and the set $S$ of existing proof trees.
Algorithm~\ref{algo:metagen} summarizes the complete procedure of \emph{MetaGen}.

\subsection*{D. Meaningless theorems}
Tree ``grafting'' can potentially introduce meaningless theorems by combining conflicting hypotheses. 
For example, 
suppose the shallow tree proves that $x^2=1$ and $x>0$ entail $x=1$, we can replace the leaf node $x>0$ with a subtree proving $x=5$ entails $x>0$, which leads to a new tree proving that $x=5$ and $x^2=1$ entail $x=1$, which is meaningless. Unfortunately, there does not appear to be an easy way to avoid meaningless theorems resulting from tree grafting, because this would require checking the consistency of an arbitrary set of expressions, which can be as hard as general theorem proving. Despite this limitation, however, we still perform tree grafting because a lot of interesting mathematics do result from nontrivial combination of hypotheses. 

\begin{algorithm}[t]
   \caption{Initializing existing proof trees}
   \label{algo:initial}
\begin{algorithmic}
\STATE {\bfseries Input:} existing theorems $E$, existing proofs $P$
\STATE {\bfseries Output:} existing proof trees $G$
    \STATE $G\leftarrow\emptyset$
    \FOR{theorem $t$ \textbf{in} $E$}
        \FOR{hypothesis $h$ \textbf{in} $\mathbf{h}_t$}
            \STATE Add $h$ to $G$
        \ENDFOR
        \STATE Add $t$ to $G$ as a one-step proof tree
    \ENDFOR
    \FOR{proof tree $p$ \textbf{in} $P$}
        \FOR{node $e$ \textbf{in} $p$}
            \STATE $g\leftarrow$ the largest subtree of $p$ rooted at $e$. 
            \STATE Add $g$ to $G$
        \ENDFOR
    \ENDFOR
\end{algorithmic}
\end{algorithm}

\begin{algorithm}[!ht]
   \caption{Constructing a proof step}
   \label{algo:step}
\begin{algorithmic}
\STATE {\bfseries Input:} existing proof trees $G$, invocable theorems $I$
\STATE {\bfseries Output:} proof step $(t,\phi)$, proof trees $S$
    \STATE Sample an invocable theorem $t\in I$
    \STATE $\phi,S\leftarrow \emptyset,\emptyset$
    \FOR{hypothesis $h$ \textbf{in} $\mathbf{h}_t$}
        \STATE $C\leftarrow\{\:g\:|\:g\in G\: \land\:$\texttt{Reachable}$(h,r_g, \phi)\:\}$
        \STATE\COMMENT{$r_g$ is the root node of proof tree $g$. $C$ is the set of compatible existing proof trees}
        \STATE Sample a proof tree $g\in C$ using softmax of the relevance network scores
        \STATE $\phi'\leftarrow$ the substitution that transforms $h$ to $r_g$
        \STATE Add $\phi', g$ to $\phi, S$
    \ENDFOR
    \FOR{variable $f$ \textbf{in} $b$}
        \IF {$f$ \textbf{not in} $\phi$}
            \STATE Generate an expression $e$ using the substitution network
            \STATE $\phi(f)\leftarrow e$
        \ENDIF
    \ENDFOR
\end{algorithmic}
\end{algorithm}

\begin{algorithm}[!ht]
   \caption{MetaGen}
   \label{algo:metagen}
\begin{algorithmic}
\STATE {\bfseries Input:} existing theorems $E$, existing proofs $P$, int $N$
\STATE {\bfseries Output:} generated theorems
\STATE Initialize existing proof trees $G$ from $E$ and $P$
\REPEAT
    \STATE Construct a proof step $(t,\phi)$ with proof trees $S$
    \STATE $g\leftarrow$ the one-step proof tree of $(t,\phi)$
    \FOR{hypothesis $h$ in $\mathbf{h}_t$}
        \STATE\COMMENT{$h(\phi)$ is a leaf node of the one-step proof tree $g$}
        \STATE Find $s\in S$ such that $r_s=h(\phi)$
        \STATE Replace $h(\phi)$ with $s$ in $g$ \COMMENT{tree grafting}
    \ENDFOR
    \STATE Add the new tree $g$ to $G$
\UNTIL{$G$ reaches the expected volume $N$}
\end{algorithmic}
\end{algorithm} 

\subsection*{E. Holophrasm}
\label{app:holophrasm}

In this section we provide more background on the Holophrasm prover~\cite{whalen2016holophrasm}. we refer the reader to ~\citet{whalen2016holophrasm} for more details. 

\noindent\textbf{Backward Reasoning}
To construct a proof tree of a target theorem, a straightforward strategy is to search backwards. We start with a single root node---the assertion of the new theorem---and pick a proof step that  establishes the entailment of the root node. We expand the tree by adding the preconditions of this proof step as children of the root node. 
We repeatedly expand the tree by adding children to leaf nodes, until each leaf node is either empty or a hypothesis of the target theorem. This construction process can be understood as recursive goal decomposition: the assertion of the target theorem is the original goal; by picking a proof step we decompose the original goal into subgoals, which are the preconditions of the proof step; then for each subgoal we repeat this process until all subgoals are resolved. 

Obviously, each time we expand the tree, we may have multiple choices of proof steps and most of them will lead to dead ends. We thus need to explore multiple alternatives, which gives rise to a search process where we need to keep track of what paths  have been explored and decide which paths to explore further.

\noindent\textbf{Proof search} Backward reasoning in Holophrasm~\cite{whalen2016holophrasm} is implemented with a proof search tree, which keeps track of the exploration of multiple branches of actions to search for a complete proof tree.
A proof search tree has two kinds of nodes, expressions and proof steps.
An expression node has multiple proof steps as children and each proof step 
establishes the entailment of this expression by the preconditions.
A proof step node has its preconditions as children.
A expression is labeled solved 
if it is a hypothesis of the target theorem or
any proof step in its children is solved. 
A proof step is labeled solved if it
has no precondition or all of its preconditions 
are solved.
A complete proof is found if the 
root node, which is the assertion of the target theorem,
is solved.

Holophrasm maintains a payoff of each node in the proof search tree and uses Monte Carlo Tree Search (MCTS) to extend the proof search tree.
The prover runs in iterations.
In each iteration, it travels down from the root node.
After visiting an expression, it either creates a new proof step as a new child 
or visits its best-performing child
according to the UCB~\citep{kocsis2006bandit} algorithm.
After visiting a proof step, 
it travels to its worst-performing child with the lowest payoff. 
When an expression node is created, 
it is assigned an initial payoff and has no children.
When a proof step node is created, 
its preconditions are also created as its children and the payoff of this proof step is the lowest payoff among its children. 
A pass continues until a new proof step is created. 

The main heuristics of the prover
are how to construct a proof step and what is the initial payoff of an expression.
Similar to the generator,
the prover constructs a proof step by using a relevance network
to pick a background theorem,
and a substitution network 
to generate a substitution for the selected background theorem.
The initial payoff of an expression is calculated by a payoff network.

\begin{table*}[t]
\caption{Training details of the relevance network and the substitution network of the prover.}
\label{tab:hyper}
\vskip 0.1in
\begin{center}
\begin{small}
\begin{sc}
\begin{tabular}{c c c c c c c}
\toprule
\multicolumn{1}{c}{\bf Network  } 
&\multicolumn{1}{c}{\bf Data  } 
&\multicolumn{1}{c}{\bf Human  } 
&\multicolumn{1}{c}{\bf Synthetic data}
&\multicolumn{1}{c}{\bf Training  } 
&\multicolumn{1}{c}{\bf Initial }
&\multicolumn{1}{c}{\bf Epoch to halve } \\
\multicolumn{1}{c}{\bf   } 
&\multicolumn{1}{c}{\bf set  }
&\multicolumn{1}{c}{\bf proofs } 
&\multicolumn{1}{c}{\bf per batch}
&\multicolumn{1}{c}{\bf  epochs } 
&\multicolumn{1}{c}{\bf learning rate}
&\multicolumn{1}{c}{\bf learning rate }
\\ \midrule
relevance & iset & 100\% & 20\% & 40 &  $10^{-3}$ & [12 20 28] \\
substitution & iset & 100\% & 70\% & 40 &  $5\times10^{-4}$ & [16, 24, 32] \\ \midrule
relevance & set & 0\% & 100\% & 5 &  $10^{-3}$ & - \\
substitution & set & 0\% & 100\% & 5 &  $5\times10^{-4}$ & - \\ \midrule
relevance & set & 10\% & 70\% & 20 &  $10^{-3}$ & [8, 12, 16] \\
substitution & set & 10\% & 70\% & 60 &  $5\times10^{-4}$ & [15, 30, 45] \\ \midrule
relevance & set & 100\% & 50\% & 16 &  $10^{-3}$ & [5, 12, 14] \\
substitution & set & 100\% & 50\% & 24 &  $5\times10^{-4}$ & [10, 15, 20] \\
\bottomrule
\end{tabular}
\end{sc}
\end{small}
\end{center}
\vskip -0.175in
\end{table*}

\paragraph{Relevance network of Holophrasm}
The relevance network of the prover is a deep network trained to pick a background theorem $b$ to establish the entailment of an expression $e$, for the purpose of proving a target theorem $t$.
It takes as input two sequences of symbols.
One sequence represents the assertion and hypotheses of $b$.
Another one represents $e$ and the hypotheses of $t$.
Two GRU encoders convert each sequence to an embedding vector,
followed by a bilinear layer
to output a score from two embeddings.
The background theorem with the highest score is selected to construct the next proof step. 
The relevance network 
is trained to pick the
background theorem that is used in the groundtruth proof step.

\paragraph{Substitution network of Holophrasm}
The substitution network 
generates the substitution for a target variable 
of a background theorem $b$ 
for the purpose of proving a target theorem $t$.
It is a sequence-to-sequence model
with an encoder-decoder GRU network.
It takes as input a sequence of symbols 
that represents
 the hypotheses of $t$
and the hypotheses of 
$b$. The target variable is replaced by 
a special token.
It is trained to generate the substitutions of groundtruth proof steps under teacher forcing.
When it is called by the prover, it 
generates multiple substitution candidates for each target variable via beam search.

\paragraph{Payoff network of Holophrasm}
The payoff network calculates the payoff of an expression as  
the probability 
of this expression being used in the proof tree of a target theorem.
It consists of a GRU network
followed by two linear layers and the sigmoid,
and takes as input a sequence of symbols that represents the 
expression to be evaluated and the hypotheses of the target theorem.

The payoff network is trained as a binary classifier to distinguish the expressions in groundtruth proof trees (called positive expressions)
from other expressions. 
Since the payoff network is used to evaluate an expression added to the proof search tree, which is a precondition of a newly generated proof step, 
the training examples of the payoff network are generated in a similar way.
For each positive expression,
 proof steps 
that establish the entailment of this expression are constructed 
by using the pretrained relevance and substitution network.
The positive expressions from the preconditions of these proof steps are filtered out  
and the payoff network is trained to distinguish the positive expressions from the rest of preconditions.

\subsection*{F. Additional Implementation details}
\label{app:detail}

We implement MetaGen and Holophrasm with the same network architectures 
as used by~\cite{whalen2016holophrasm}.
For all of our networks in the generator and the prover, we use bidirectional GRUs to encode input sequences,
and use the Adam~\citep{kingma2014adam} optimizer to update parameters.
The batch size is 100 unless otherwise noted.

\noindent\textbf{Task setup}
It is important to note that a theorem can serve both as a target theorem in the test set and as a background theorem in the training set. This is a standard setup and is not ``training on the test set''---a background theorem is used as a known fact in a training proof task and only its statement is provided, not its proof; seeing the statement of a background theorem during training does not tell us how to prove it during testing. 

\noindent\textbf{Input representation of the relevance and substitution network}
Here we provide more details on the input representation of the relevance and substitution network, which take sequences as input. 
We use the same form of input representations 
 as used by~\citet{whalen2016holophrasm}.

To represent an expression in a sequential form, one option is to use its ``surface form''.  
For example, ``(1+1)=2'' is simply given as such. Another option is to serialize its parse tree. 
The parse tree of ``(1+1)=2'' has two generating axioms.
The first axiom is the root node of its parse tree and generates an expression in the form of ``A=B''.
The second axiom is the left child of the root node and generates 
an expression in the form of ``(C+D)'' and this expression is used to substitute the variable A in the first axiom.
The right child of the first axiom is the token ``2''.
Both of the left child and the right child of the second axiom are the token ``1''. 
Then we can represent ``(1+1)=2'' as a sequence of symbols
$(t_=, t_+, 1, 1, 2)$,
where each symbol is a node in the parse tree and $t_=$ and $t_+$ represent two generating axioms.
This new sequence is obtained by traversing the parse tree in pre-order. Following ~\citet{whalen2016holophrasm}, we use the second option to represent expressions as input to our network.

Following~\citet{whalen2016holophrasm},
we also make use of the graph structure of the parse tree.
Each node in the input sequence is converted 
to a feature vector by a learnable embedding layer. 
Then the feature of this node is concatenated with another four-dimension vector describing  
the depth of the node, the degree of the node, the degree of its parent, and its position into the children of its parent.
The concatenated vector is fed into the GRU encoder of the relevance and substitution network.

Multiple expressions are represented by 
their concatenation.

\begin{table*}[t]
\caption{Examples of synthetic theorems from \textit{MetaGen-IL} trained on all human proofs of \texttt{set.mm}.}
\label{table:examples}
%\vskip 0.15in
\begin{center}
\begin{small}
\begin{sc}
\begin{tabular}{c c c}
\toprule
\multicolumn{1}{c}{\bf  Hypothesis } 
&\multicolumn{1}{c}{\bf Assertion } 
&\multicolumn{1}{c}{\bf Comment}
\\ \midrule
$\emptyset$ & $( 3 \times 1 ) + ( 1 + 0 ) = 1 + 3$ & Simple arithmetic.\\ 
\midrule
$\emptyset$& $( \log e ) \times A  = A$ & $e=2.71828...$\\ 
\midrule
$A \in \mathbb{C}$ & $\sin ( A + B ) =(\exp (\mathbf{i}\times ( A + B) )$ & $\mathbb{C}\colon$ complex number set. \\
$B \in \mathbb{C}$ & $ - \exp ( - \mathbf{i}  \times ( A + B )) \div (2 \times \mathbf{i})$ & $\mathbf{i} = \sqrt{-1}$. \\  \midrule
$\emptyset$ & $ G \in \mathbb{R} \land E \in \mathbb{R} \rightarrow \sin ( \frac{ G + E } { 2 } + 1) \in \mathbb{R}$ & $\mathbb{R}\colon$  real number set.\\ \midrule
$\phi \rightarrow F \colon X\leftrightarrow Y$  & 
$\phi \rightarrow \text{Ran} (F) \subseteq Y$ 
&F: bijection from X to Y. \\
&& $\text{Ran}(F)$: range of $F$.\\ \midrule
$N = \{ x \in \mathbb{Z} | M \leq x \}$ & $\phi\land K \in N\rightarrow$  & $\mathbb{Z}\colon$ integer set \\ 
&$ M \in \{ x \in \mathbb{Z} | M \leq x \land x \leq K \}$ &\\
\midrule
$r =  q \times 2 \times y \mod p$ &$x = y \rightarrow F ( r \times y ) = F ( s \times x ) $ & mod: modulo operation \\
$s = q \times 2 \times x \mod p$  && \\ 
\bottomrule
\end{tabular}
\end{sc}
\end{small}
\end{center}
%\vskip -0.1in
\end{table*}

\subsubsection*{F.1. Generator}
\label{app:generator}
\noindent\textbf{Configuration of GRUs} All of the GRUs in the generator have two layers and 256-dimensional hidden units. 

\noindent\textbf{Training relevance network of \emph{MetaGen-IL}}
The relevance network of \emph{MetaGen-IL} is updated to minimize the cross-entropy loss. Each training sample has one groundtruth proof tree and 10 negative candidates that are randomly sampled
from compatible proof trees.
It is trained for 60 epochs.
The learning rate is set to $10^{-4}$ initially and halved
after 30, 40 and 50 epochs. 

\noindent\textbf{Training substitution network of \emph{MetaGen-IL}} The substitution network of \emph{MetaGen-IL} is trained for 40 epochs.
The learning rate is set to $5\times10^{-4}$ initially and halved 
after 20, 26 and 32 epochs. 

\noindent\textbf{Training of \emph{MetaGen-RL}}
To train \emph{MetaGen-RL-LM},
we learn the language model of human-written theorems 
by utilizing
a one-layer GRU with 64-dimensional hidden units. 
It is trained for 200 epochs.
The learning rate is set to $5\times10^{-4}$ initially and halved after 80, 120 and 160 epochs. 

To train \emph{MetaGen-RL-Adv}, 
we train a binary classifier using the same architecture 
as the payoff network of Holophrasm, 
which contains a two-layer GRU with 128-dimensional hidden units and two subsequent linear layers.
It is pretrained to distinguish human-written theorems from 300K synthetic theorems generated by \emph{MetaGen-Rand}.
Then it is updated on-the-fly to distinguish human-written theorems from the synthetic theorems generated in the most recent 20 episodes. 

For both \emph{MetaGen-RL-LM} and \emph{MetaGen-RL-Adv},
we train the generator 
for 700 episodes with the learning rate fixed to $10^{-4}$.
We deploy 10 parallel threads to synthesize new theorems by utilizing the current generator. Each thread generates 50 theorems 
in one episode and synchronizes the set $G$ of existing proof trees with other threads for every 20 episodes. 
We clip policy gradients whose norm is larger 5. 

\subsubsection*{F.2. Prover}
\label{app:prover}
\noindent\textbf{Configuration of GRUs}
In the relevance and substitution network of the prover, 
all GRUs have two layers and 256-dimensional hidden units.
We found 256-dimensional GRUs have slightly better performance than the 128-dimensional GRUs that are used by~\citet{whalen2016holophrasm}.
The GRU in the payoff network of the prover
has two layers and 128-dimensional hidden units.

\noindent\textbf{Training of the prover}
All three networks of the prover are trained by imitation  learning. 
The relevance network and the substitution network
are trained on both human-written proofs and  synthetic proofs.
The payoff network is trained on human-written proofs only.

The relevance network of the prover 
is trained to minimize the cross-entropy loss.
Each training sample contains 
one groundtruth background theorem
and 10 negative candidates that are randomly sampled
from all background theorems that can be applied in this step.

Table~\ref{tab:hyper} presents the settings of learning rate schedules 
and
the ratio of synthetic training samples per batch, 
for the training of the relevance and substitution network
of the prover.

In all experiments, the payoff network is trained for 30 epochs. 
The learning rate is set to $10^{-4}$ initially and halved
after 15, 20 and 25 epochs.

\noindent\textbf{Evaluation protocol}
Following the evaluation protocol used by~\citet{whalen2016holophrasm},
the prover attempts to prove each target theorem in the test set
three times with the beam search width of the substitution network set to 1, 5, or 20.
The prover stops if it has executed 10000 MCTS passes
or hit the time limit of 5 minutes.

\subsubsection*{F.3. Baseline}

Without human-written proofs,
we compare our approach with a baseline that needs no training proofs.
We remove the relevance network of the prover 
and pick a background theorem according to 
the tf-idf similarity between an expression and a background theorem, as proposed by~\citet{bansal2019learning}.
We replace the substitution network of the prover 
with a language model trained on the statements of human-written theorems.
We use this language model to generate an expression as the substitution of a target variable.

\subsection*{G. Examples of generated theorems}
Some examples of synthetic theorems are presented in the Table~\ref{table:examples}. Some are trivial (first and fourth), whereas others are fairly interesting---the third theorem involves a non-trivial statement about trigonometric functions and complex numbers. 

\end{document}